%% file: main.tex
\documentclass{article}
\input{header}
\input{makros}

\title{Comparing the 3D morphology of solid-oxide fuel cell anodes for different manufacturing processes, annealing times, and operating temperatures }
\author{Sabrina Weber $^{1,\ast}$, Benedikt Prifling $^{1}$,  Martin Juckel$^{2}$, Yanting Liu$^3$, Matthias Wieler$^4$, 
Daniel Schneider$^5$, Britta Nestler$^5$, Norbert H. Menzler$^2$, Volker Schmidt$^{1}$\\}

\date{}

\begin{document}

\maketitle
\vspace{-4em}
\begin{center}
	\it
	$^1$ Institute of Stochastics, Ulm University, 89069 Ulm, Germany\\
	$^2$  Institute of Energy Materials and Devices, 
 IMD-2: Materials Synthesis and Processing, 52428 Jülich, Germany
	\\
    $^3$ Institute for Applied Materials (IAM-ET), Karlsruhe Institute of Technology (KIT), 76131 Karlsruhe, Germany\\
	$^4$ Robert Bosch GmbH (BOSCH), Zentrum für Forschung und Vorausentwicklung, 71272 Renningen, Germany\\
    $^5$ Institute for Applied Materials (IAM-MMS), Karlsruhe Institute of Technology (KIT), 76131 Karlsruhe, Germany\\
 	
\end{center}
\blfootnote{$^\ast$Corresponding author \\ \textit{Email addresses: sabrina.weber@uni-ulm.de}}

\input{abstract}
\section{Introduction}

Solid oxide fuel cells (SOFCs) are becoming increasingly important due to their high electrical efficiency, the flexible choice of fuels and relatively low emissions of pollutants \cite{weber2021fuel,kendall.2015, fergus.2016, brandon.2017}. Thus, applications of SOFCs range from stationary power generation in the private as well as the industrial sector to electric vehicles, where they can be considered as emerging technology. However, the steadily growing demands for electrochemical devices in combination with open problems such as pronounced degradation mechanisms require further technological progress to unlock the full potential of fuel cells as an efficient alternative to traditional fossil fuel-based energy sources\cite{singh.2021, zarabi.2022}. One possibility to achieve this goal is the transition towards novel materials as for example the replacement of yttria-stabilized zirconia (YSZ) by gadolinium-doped ceria (GDC), which offers a higher ionic conductivity at lower temperatures compared to YSZ \cite{zhang2020comparison, chueh.2012, nenning.2020, hussain.2020}. This also reduces thermal degradation effects and thus increases the durability of SOFCs. Besides further optimizing the performance of SOFCs via improving the properties of the underlying materials, tuning the 3D morphology  of the electrodes is a promising strategy since it is well known that the 3D morphology  has a profound impact on fuel   transport  within the pore space, electric transport as well as ionic transport within the electrolyte \cite{szasz2015nature, suzuki.2009, connor.2018, pecho.2015}. Moreover, not only the spatial distribution of each phase is of interest, but also the length of the triple-phase boundary (TPB), where the chemical reaction mainly takes place. More precisely, at the anode the fuel gas (typical choices are hydrogen or methane) is oxidized, which results in electrons that are transported to the cathode through an external circuit. At the cathode, these electrons are used to power the reduction reaction of oxygen, which react to oxygen ions. In order to quantitatively characterize the 3D morphology  of  SOFC electrodes at the relevant scale, 
focused ion beam scanning electron microscopy
(FIB-SEM) is a frequent choice \cite{wilson.2006, vivet.2011, kishimoto.2016, taillon.2014, meffert.2020}. \\

Note that the 3D morphology of GDC-based SOFC anodes significantly differs from that of the conventional YSZ-based anodes. In \cite{nenning.2020}, it is shown that the low polarization resistance of GDC-based anodes is related to the material properties and the microstructure of the anode. Additionally, several studies have examined the degradation effects on microstructure. For example, the studies in \cite{holzer2011microstructure, holzer2011quantitative} investigate degradation effects for long term exposition to humidified and dry hydrogen atmospheres, analyzing volume fractions,  particle size distributions,  TPB and specific surface areas. Similarly, degradation effects on the microstructure for cell operation up to 20000 hours are analyzed in \cite{zekri2017microstructure} by computing geometrical descriptors such as TPB, particle size distribution and porosity, showing that the nickel and GDC phases tend to coarsening. A shorter operational period is considered in \cite{sciazko2020multiscale} to analyze the changes in the microstructure of nickel and GDC grains as well as the overall microstructure. Moreover, it turned out that local heterogeneities of electrodes have a profound impact on the  resulting performance~\cite{mahub.2021, epting.2017,hsu.2018}.\\

In the present paper, we consider eight GDC-based SOFC anodes, which are imaged via 3D FIB-SEM. The resulting image data is used as a basis for a comprehensive structural characterization of these anodes, which differ with regard to their operating conditions (\ie operating temperature and annealing time) and the underlying manufacturing process (powder technology or infiltration). The quantitative structural analysis of these SOFC anodes by means of various geometrical descriptors allows for a deeper understanding of process-structure relationships and degradation phenomena, which is crucial for the design of SOFC anodes with improved electrochemical properties. \\

The rest  of this paper is organized as follows. In Section~\ref{sec:experimental}, the manufacturing process of the SOFC anodes including the underlying materials and the operating conditions, the imaging via 3D FIB-SEM  and the segmentation of the resulting image data into nickel, \cgo and pore space is explained. Next, the geometrical descriptors that are used for the quantitative structural analysis of the anodes - including, among others, volume fraction, specific surface area, and mean length of TPB per unit volume - are presented in Section~\ref{sec: geom_desc}. Section~\ref{sec:results} contains a detailed, quantitative analysis of the influence of  the operating temperature,  the annealing time, as well as the manufacturing process on the resulting 3D morphology of the anodes. Finally, the paper is concluded by a summary of the main results and an outlook to future research activities.

\section{Experimental}\label{sec:experimental}

This section covers the description of the manufacturing processes for seven SOFC anodes, including  underlying materials as well as  operating conditions (Section~\ref{subsec:materials}),  tomographic imaging via 3D FIB-SEM (Section~\ref{subsec:tomographic_imaging}) and, finally,  segmentation of the 3D image data into nickel, \cgo and pore space  (Section~\ref{subsec:imageprocessing}). 

\subsection{Electrode materials, manufacturing and aging}\label{subsec:materials}

In the present paper, two different manufacturing processes for SOFC anodes, which have an electrode surface area of \SI{1}{cm^2}, are considered:  powder technology and infiltration.\\ 

In the first case, symmetrical cells consisting of an 8 mol-\% YSZ electrolyte from the company Kerafol\textsuperscript{\textcopyright} (Eschenbach i.d.Opf., Germany) sandwiched in between \cgo interlayers were produced. The \cgo powder ($\mathsf{Gd}_{0.1}\mathsf{Ce}_{0.9}\mathsf{O}_{1.95}$) was commercially available from Fuelcellmaterials and transformed into a paste by the standard procedure of Forschungszentrum J\"{u}lich GmbH, using $\alpha$-terpineole as a dispersion medium and ethylcellulose as a transport medium. The GDC interlayer was sintered at \SI{1300}{\celsius} for three hours.
The functional layer (50 w\% NiO, 50 w\% GDC) was produced in the same way using, commercially available \cgo powder ($\mathsf{Gd}_{0.1}\mathsf{Ce}_{0.9}\mathsf{O}_{1.95}$) from Fuelcellmaterials and $\mathsf{NiO}$ powder from G. Vogler B.V. in a 50:50 w\% ratio. The functional layer was sintered at \SI{1400}{\celsius} for three hours.\\

For the infiltration experiment, an additional $\mathsf{NiO}$ layer was screen printed on both sides of the \cgo sandwiched 8 mol-\% YSZ electrolyte. Note that a NiO layer is chosen instead of a pure Ni layer due to facilitate the infiltration process. The infiltration of the $\mathsf{NiO}$ layer was conducted, using a $\mathsf{\cgo}(\mathsf{NO}_{3})_{3}$ solution ($\SI{3}{\mole}$). Therefore, the cells were immersed into the solution for about five minutes under vacuum, dried and sintered at $\SI{500}{\celsius}$ for $\SI{3}{\hour}$. Different sintering temperatures for the scaffold were used and a variety of repetitions for the infiltration steps was tested. In the above-described cases, an additional contact layer of $\mathsf{NiO}$ was added on top, using screen printing method and dried at $\SI{80}{\celsius}$ overnight.\\

In case of the SOFC anodes manufactured via the infiltration process, the cells were gradually heated in a mixture of \SI{95}{\percent} $\mathsf{N}_{2}$ and \SI{5}{\percent} $\mathsf{H}_{2}$ at a rate of $\SI{1}{\kelvin\per \minute}$ until reaching $\SI{650}{\celsius}$. At this temperature, the gas composition was changed to \SI{65}{\percent} $\mathsf{N}_{2}$, $\SI{30}{\percent} \mathsf{H}_{2}$, and $\SI{5}{\percent} \mathsf{H}_{2}\mathsf{O}$ and maintained for six hours. Subsequently, the gas composition was adjusted to $\SI{3.7e-4}{\mole\per\second}$ of hydrogen and  $\SI{9.2e-5}{\mole\per\second}$ of oxygen to create a 1:1 hydrogen/steam mixture. Electrochemical impedance spectroscopy (EIS) was then used to characterize the cells at $\SI{650}{\celsius}$ and $\SI{600}{\celsius}$, see \cite{liu.2023} for further details. Two of the three cells, which are manufactured via the infiltration process, were subjected to additional high-temperature treatments under the aging conditions detailed in Table~2 of \cite{liu.2024}. During the aging process, reference impedance measurements at $\SI{600}{\celsius}$ were conducted on both cells every few hundred hours. 

\subsection{Tomographic imaging}
\label{subsec:tomographic_imaging}

Tomographic imaging has been carried out with 3D FIB-SEM using a Helios 5 Hydra DualBeam (ThermoFisher) with a concentric backscatter detector. An acceleration voltage of $\SI{3}{\kilo\eV}$ and a current of $\SI{0.8}{\nano\ampere}$ is used. Some samples have been imaged with a voxel size of $\SI{20}{\nano\meter}$ and some with $\SI{50}{\nano\meter}$, see Table~\ref{tab:my_label} for a detailed overview of all considered anode samples. Note that for anodes manufactured by means of the infiltration technology a higher resolution is chosen, due to the finer microstructure. For anodes manufactured by means of the powder technology a lower resolution is sufficient for the microstructure analysis. However, to compare the two manufacturing processes for sample G, a resolution of \SI{20}{nm} is required.

\begin{table}[htb]
    \centering
    \begin{tabular}{p{2.5em}|p{6.7em}|p{7em}|p{6.5em}|p{7em}|p{6.2em}|p{8em}}
         sample & manufacturing process & reduction temp. [\SI{}{\celsius}] & annealing time [\SI{}{\hour}] & operating temp.  [\SI{}{\celsius}] & voxel size [\SI{}{\nano \meter}] & image size [\SI{}{\micro \meter}] \\
         \toprule
         A & powder & 800 & 0 & - & 50 & $6.4\times 37.0 \times 28.2$ \\
         B & powder & 800 & 240 & 900 & 50 & $8.0 \times 25.0 \times 24.0$\\
         C & powder & 800 & 1100 & 900 & 50 & $7.7 \times 22.7 \times29.2 $\\
         D & infiltration & 650 & 640 &  900 & 20 & $4.5 \times 17.1 \times 14.2$ \\
         E & infiltration & 650 & 0 & - & 20 & $7.9 \times 15.8 \times 13.6$ \\
         F & infiltration & 650 & 1000 & 700 & 20 & $5.9 \times 19.7 \times 17.6$ \\
         G & powder & 650 & 0 & - & 20 & $6.9 \times 18.8 \times 20.6 $ \\
    \end{tabular}
    \caption{Sample information including manufacturing process, reduction temperature, annealing time, operating temperature and voxel size.}
    \label{tab:my_label}
\end{table}

\subsection{Image processing}\label{subsec:imageprocessing}

At first, the 3D grayscale image data has been denoised using a total variation filter with a denoising weight of 50 \cite{getreuer.2012}. Next, the grayvalue histogram of the denoised 3D image is computed and partitioned into five regions, namely pore markers, nickel markers, \cgo markers and two intermediate regions. Each of the two intermediate regions corresponds to 10\% of all voxels with grayvalues, where one can not yet distinguish between pore space and nickel or between nickel and \cgo, respectively. Next, markers are removed if they are close to some interface, which is quantified by the upper 20 \% quantile of highest gradient values. Note that the edge magnitude using the Scharr transform is exploited to compute the gradient \cite{kroon.2009}. Moreover, markers are removed near ``cracks'', which is defined as highest 2 \% (in case of the samples with a voxel size of $\SI{20}{\nano\meter}$) or 5 \% (in case of the samples with a voxel size of $\SI{50}{\nano\meter}$) values after Meijering filtering \cite{meijering.2004}. The remaining markers are then used as markers for the watershed algorithm, where the input image is given by the gradient image \cite{beucher.1993}. 

\section{Geometrical descriptors}
\label{sec: geom_desc}

In order to quantify the 3D morphology of the SOFC anodes considered in this paper, various geometrical descriptors are used. For this,  the three phases, \ie nickel, \cgo and pores, are  considered as  realizations of  random closed sets (RACSs) in the three-dimensional Euclidean space $\mathbb{R}^3$. These RACSs are assumed  to be motion-invariant, \ie  stationary and isotropic, 
and denoted by 
$\Xi_1, \Xi_2$ and $\Xi_3$,
respectively.
Moreover, in the definition of some geometrical descriptors, we even assume that the vector $(\Xi_i,\Xi_j)$ is motion-invariant for any $i,j\in\{1,2,3\}$. On the other hand, in some cases, the assumption of stationarity is sufficient.
 A formal introduction to the theory of stationary and isotropic RACSs can be found in \cite{molchanov.2005}.

\paragraph{Volume fraction} One of the most fundamental geometrical descriptors of stationary RACSs is their volume fraction. Furthermore, the volume fractions of the three anode phases  mentioned above  have a profound impact on the performance of SOFCs. In the following, the volume fraction of a stationary RACS  will be denoted by $\varepsilon$. It can be easily estimated from voxelized 3D image data via the point-count method \cite{chiu.2013}.

\paragraph{Specific surface area} Similar to  volume fractions, the specific surface areas (SSAs)  of the three anode phases are also of importance with regard to the performance of SOFCs \cite{brandon.2017}, where this geometrical descriptor is defined as the mean surface area of a predefined phase per unit volume.  It will be denoted by $S$ in the following. For the estimation of $S$ from voxelized 3D image data, locally weighted $ 2\times 2\times 2$ voxel configurations are used, see \cite{schladitz.2007} for details regarding the choice of the weights. Thus, each of the $2^{8}=256$ possible configurations of $ 2\times 2\times 2$ voxels belongs to one of fourteen classes, which have different weights. The surface area is then obtained by iterating over the complete 3D image and adding up the corresponding weights. Besides the SSA  of a certain phase, the specific interfacial area between two phases is also of interest. Note that in case of the three-phase anode  material, which is modeled by the stationary  RACSs $\Xi_{1}, \Xi_{2}$ and $\Xi_{3}$, the specific interfacial area $ S(\Xi_{i} \cap \Xi_{j})$ between phases $i$ and $j$ with $i,j\in \{1,2,3\}$ and $i\neq j$ can be determined via the following equation:
\begin{equation}
    S(\Xi_{i} \cap \Xi_{j}) =\frac{1}{2}\Bigl(S(\Xi_{i}) + S(\Xi_{j}) - S(\Xi_{k})\Bigr),
\end{equation}
where $k = \{1,2,3\} \backslash \{i,j\}$. As highlighted in \cite{kishimoto.2016}, the GDC pore interface is particularly important as an increased GDC pore interface increases performance by providing more area for electrochemical activity.

\paragraph{Triple-phase boundary length} Since the chemical reaction in SOFC anodes takes place at the triple-phase boundary (TPB), the expected length of TPB per unit volume is of particular interest. It is called specific length of TPB in the following and defined as
\begin{equation*}
    \ell_{\mathsf{TPB}} = \mathbb{E}\left( \mathcal{H}_{1}
(\Xi_1 \cap \Xi_2 \cap \Xi_3 
\cap [0,1]^{3}) \right),
\end{equation*}
where $\mathcal{H}_{1}$ denotes the one-dimensional Hausdorff measure in $\mathbb{R}^3$. This geometrical descriptor is estimated from discretized 3D image data by first detecting all $2\times 2 \times 2$ voxel configurations, which contain at least one voxel of each of the three phases. Next, each pair of these voxel configurations that are neighbors with respect to the 6-neighborhood is determined. Finally, the number of these pairs is divided by the total number of voxels in the sampling window. Note that there are various approaches for estimating the length of the TPB from voxelized image data \cite{wilson.2010, iwai.2010, jorgensen.2014, jorgensen.2010}.

\paragraph{Geodesic tortuosity} In SOFC anodes, three kinds of transport processes take place: The transport of the gaseous educts and products within the pore space, the transport of oxygen ions within the \cgo phase, and the transport of electrons, which mainly takes place within the nickel phase. A geometrical descriptor that turned out useful to characterize various types of transport phenomena in porous media is connected with the tortuosity, \ie the windedness of their phases. Note that there exist several  notions of tortuosity  in the literature \cite{holzer.2023}. In the present paper, we consider the so-called geodesic tortuosity, which is a purely geometric descriptor of 3D morphologies. More precisely, a starting plane and a target plane, which is parallel to the starting plane, are chosen. Next, for a given phase of the material, the shortest path to the target plane is determined for each point of this phase within the starting plane. Then, the lengths of these shortest paths are normalized by dividing them  by the distance between both planes. This results in the distribution of geodesic tortuosity, the mean value of which will be denoted by $\tau$. A formal definition of geodesic tortuosity within the framework of random closed sets is presented in \cite{neumann.2019}. The computation of this geometrical descriptor from voxelized 3D image data is carried out by means of Dijkstra's algorithm \cite{jungnickel.2007}. \\

Moreover, two modifications of  geodesic tortuosity  are considered. On the one hand, we consider the so-called dilated geodesic tortuosity, where we just investigate the lengths of those paths along which there is a certain minimum distance $r>0$ to the complement of the given transport phase. On the other hand, only paths to the target plane are considered which start from the TPB, since the majority of transport processes are either targeting the TPB or starting at the TPB. 

\paragraph{Constrictivity} The constrictivity of a transport phase, denoted by $\beta \in [0,1]$, is a measure for the strength of bottleneck effects. It is defined as $\beta = \left( r_{\mathsf{min}} \slash r_{\mathsf{max}} \right)^{2}$, where $\beta=1$ corresponds to the case that no bottleneck effects exist, and a value of $\beta$ close to zero indicates pronounced bottleneck effects \cite{muench.2008}. The quantity  $r_{\mathsf{max}}$ appearing in this formula is related with the continuous pore size distribution function $\mathsf{CPSD}: [0,\infty) \to [0,1]$ of a stationary RACS $\Xi\subset\mathbb{R}^3$ 
\cite{serra.1982, soille.2003}.  Here, for each $r\geq 0$, the subset
$\Xi \circ B(o,r)$
of $\Xi$ is considered, which can be covered by (potentially overlapping) spheres with radius $r$,  where $B(o,r)\subset\mathbb{R}^3$
is the sphere with midpoint at the origin $o\in\mathbb{R}^3$ and radius $r\ge 0$, and
$\circ$ denotes  morphological opening. The value $\mathsf{CPSD}(r)$  is then defined as the volume fraction of the stationary RACS $\Xi \circ B(o,r)$, and $r_{\mathsf{max}}=\sup\{r \geq 0: \mathsf{CPSD}(r)\ge 0.5\ \mathsf{CPSD}(0)\}$. Note that  $0.5\cdot \mathsf{CPSD}(0)$ equals half the volume fraction of the phase of interest, which will be also used for defining the characteristic bottleneck radius $r_{\mathsf{min}}$. More precisely, this geometrical descriptor is based on the simulated mercury intrusion porosimetry $\mathsf{SMIP}: [0,\infty) \to [0,1]$, which is closely related to  CPSD, but additionally depends on a predefined direction. More precisely, the value $\mathsf{SMIP}(r)$ is defined as the volume fraction of the subset of the $\Xi$, which can be reached by an intrusion of spheres with radius $r$ from the predefined direction. Analogously to $r_{\mathsf{max}}$, the radius $r_{\mathsf{min}}$ is defined as $r_{\mathsf{min}}=\sup\{r \geq 0: \mathsf{SMIP}(r)\ge 0.5\ \mathsf{CPSD}(0)\}$. In \cite{prifling.2021, neumann.2020}, it has been shown that constrictivity is a useful geometrical descriptor for characterizing transport processes, which highlights the importance of this quantity with regard to the various kinds of transport processes in SOFC anodes.

\paragraph{Two-point coverage probability function} Another frequently used geometrical descriptor is the two-point coverage probability function. More precisely, let $\Xi_{i}$ and $\Xi_{j}$ be two RACSs such that the vector $(\Xi_i,\Xi_j)$ is motion-invariant, for all $i,j\in \{1,2,3\}$. The two-point coverage probability function $\tilde{C}_{i,j}:[0,\infty) \to [0,1]$ is then defined by $\tilde{C}_{i,j}(r) = \mathbb{P}(o \in \Xi_{i}, h \in \Xi_{j})$, where $h \in \R^{3}$ is an arbitrary vector of length $r\geq 0$. In particular, the value of $\tilde{C}_{i,j}(r)$ only depends on the distance $r\ge 0$ between two points due to the assumed motion-invariance of $(\Xi_i,\Xi_j)$. Note that the  limit of $\tilde{C}_{i,j}(r)$ for $r\to\infty$  is often equal to the product $\varepsilon_{i}\varepsilon_{j}$ of the volume fractions $\varepsilon_i$ and $\varepsilon_j$ of $\Xi_i$ and $\Xi_j$, respectively,  since in case of the vast majority of materials the events that $o \in \Xi_{j}$ and $h \in \Xi_{j}$ are becoming stochastically independent for unboundedly increasing distances $r$. Thus, in the following, we  consider the centered two-point coverage probability function $C_{i,j}:[0,\infty)\to [0,1]$, which is defined as $C_{i,j}(r)  = \tilde{C}_{i,j}(r) - \varepsilon_{i} \varepsilon_{j}$ for each $r\ge 0$. \ie there is a positive correlation between the events $o \in \Xi_{i}$ and $h \in \Xi_{j}$ if $C_{i,j}(r)>0$, and vice versa. The estimation of the values of $C_{i,j}$ from voxelized image data is carried out via a Fourier-based method described in \cite{ohser.2009}. Note that in \cite{prifling.2021}, it has been shown that this geometrical descriptor can improve the prediction of effective transport properties, which indicates that this quantity is closely related to the performance of SOFC anodes. 

\paragraph{Chord length distribution} The chord length distribution of a motion-invariant RACS $\Xi$ is defined as the distribution of the length of   segments chosen at random in $\Xi \cap \ell$, where $\ell$ is some line passing through the origin Due to the  isotropy of $\Xi$, the chord length distribution does not depend on the specific choice of the direction of the intersecting line $\ell$. Thus, in the following, only chords in the direction of the $x$-axis are considered. To estimate the chord length distribution from voxelized image data, we use the algorithms proposed in  \cite{ohser.2000, ohser.2009}. Note that a shorter chord length indicates a finer phase microstructure and thus, tends to lead to more active reaction sites. 

\paragraph{Spherical contact distance} The cumulative distribution function $H:[0,\infty) \to [0,1]$ of spherical contact distances of a stationary RACS $\Xi$ is defined as
\begin{equation*}
	H(r) = 1 - \mathbb{P} (o \notin \Xi \oplus B(o,r) \mid o \notin \Xi) \qquad\text{ for each }r \geq 0,
	\label{eq:racs:scd}
\end{equation*}
where $\oplus$ denotes  Minkowski addition, \ie the value of $H(r)$ is the probability that the distance from a randomly chosen point of  the complement of $\Xi$ to the nearest point of $\Xi$ is not longer than $r$. Thus, this geometrical descriptor can be used in order to quantify the compactness/perforation of a geometrically complex phase. When estimating the spherical contact distribution from voxelized 3D image data, the Euclidean distance transform is computed and the algorithm described in \cite{mayer.2004} is applied. Note that it has been shown in \cite{prifling.2023}, that a combination of the mean spherical contact distance and the mean chord length is well suited for prediction of diffusive transport properties, which is also to be expected for SOFC anodes.

\paragraph{Local geometrical descriptors} Besides computing the geometrical descriptors mentioned above, just once for each of the entire 3D images corresponding to the eight SOFC anode samples considered in this paper, we additionally compute some of these descriptors locally to quantify the heterogeneity of the SOFC anodes. More precisely, the 3D image data is partitioned into non-overlapping cutouts. For this purpose, we consider cubic cutouts with a side length of \SI{2.5}{\micro \meter} in each direction. Thus, for a  voxel size of \SI{20}{\nano \meter} and \SI{50}{\nano \meter}, we get cubic cutouts with a size of $125$ and $50$ voxels, respectively.

\section{Results and discussion}\label{sec:results}

We now use the geometrical descriptors stated in Section~\ref{sec: geom_desc} in order to investigate the influence of  operating temperature,  annealing time and the manufacturing process on the 3D morphology of SOFC anodes. Therefore, subsets of samples from the eight SOFC anode samples considered in this paper, which differ in exactly one of these influencing factors, are quantitatively compared to each other with respect to the geometrical descriptors of Section~\ref{sec: geom_desc}. Note that in the figures presented below, results regarding the nickel phase are always shown in blue, whereas those regarding the \cgo phase are shown in red, and those regarding  the pore space in yellow.

\subsection{Influence of operating temperature}

\begin{figure}[H]
    \centering
    \includegraphics[width = 0.9\textwidth]{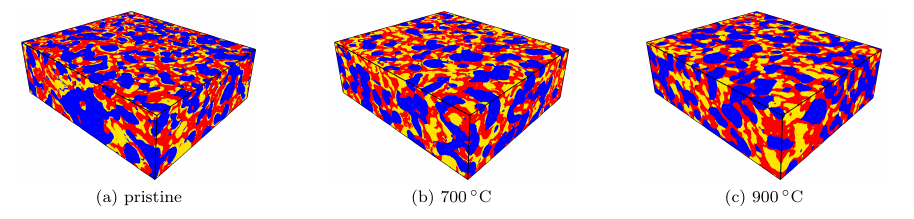}
    \caption{3D morphology of  SOFC anodes, consisting of nickel (blue), \cgo (red) and pore space (yellow), for the pristine state (a) as well as for operating temperatures of  $\SI{700}{\celsius}$ (b) and  $\SI{900}{\celsius}$ (c). Both samples are manufactured by means of the infiltration process, reduced at a temperature of \SI{650}{\celsius}, and imaged with a voxel size of $\SI{20}{\nano \meter}$.}
    \label{fig: temp_sample}
\end{figure}

To begin with, we investigate the influence of operating temperature on the 3D morphology of  SOFC anodes, where we consider samples F and D, \ie samples manufactured by means of the infiltration process, a reduction temperature of \SI{650}{\celsius} and a voxel size of $\SI{20}{\nano\meter}$.  Under these conditions, we compare the samples with an operating temperature of \SI{700}{\celsius} and \SI{900}{\celsius}, respectively, where the 3D  morphology of these samples and of the pristine state  (sample E) is  visualized in Figure~\ref{fig: temp_sample}. Note that the sample visualizations in this paper are always sample cutouts with a size of $\SI{4.4}{\micro \meter} \times \SI{11.6}{\micro \meter} \times \SI{13.6}{\micro \meter}$.

\begin{figure}[H]
    \centering
    \includegraphics[width=0.4\linewidth]{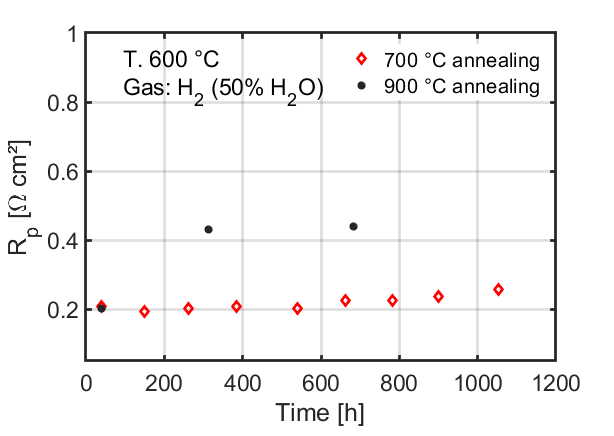}
    \caption{Time evolution of polarization resistance of the electrode of cells F and D, i.e., annealed at \SI{700}{\celsius} and \SI{900}{\celsius}, measured at \SI{600}{\celsius} under a gas mixture of steam: hydrogen 50 : 50 for reference. }
    \label{fig:polarization}
\end{figure}

However, the annealing time also differs for these two samples. For the sample with an operating temperature of \SI{700}{\celsius} the annealing time is \SI{1000}{\hour}, while the annealing time of the sample with an operating temperature of \SI{900}{\celsius} is only \SI{640}{\hour}. Nonetheless, the comparison of these two samples is reasonable considering the polarization resistance. The polarization resistances measured at 600 °C for the reference infiltrated cells annealed at 700 °C and 900 °C are plotted against time in Figure \ref{fig:polarization}. The initial values of polarization resistance ($R_{p}$)  for both cells are nearly identical, indicating good reproducibility within this batch. Over time, the polarization resistance increased by \SI{0.0574}{\ohm cm^2}, corresponding to a \SI{28}{\percent} increase after 1100 hours at \SI{700}{\celsius}. In contrast, for the cell aged at \SI{900}{\celsius}, a clear thermally activated aging effect was observed. After 310 hours, the value of $R_{p}$ reached \SI{0.431}{\ohm cm^2}, reflecting a \SI{116}{\percent} increase. Over the subsequent 330 hours, the increase in polarization resistance was much smaller, rising by only \SI{0.008}{\ohm cm^2}, which corresponds to a \SI{4}{\percent} increase from the initial value. Thus, comparing the microstructure of these two cells is reasonable, as the polarization resistance measured at 900 °C can be considered to have undergone negligible structural changes after 310 hours.

\begin{figure}[H]
    \begin{center}\includegraphics[width = 0.75\textwidth]{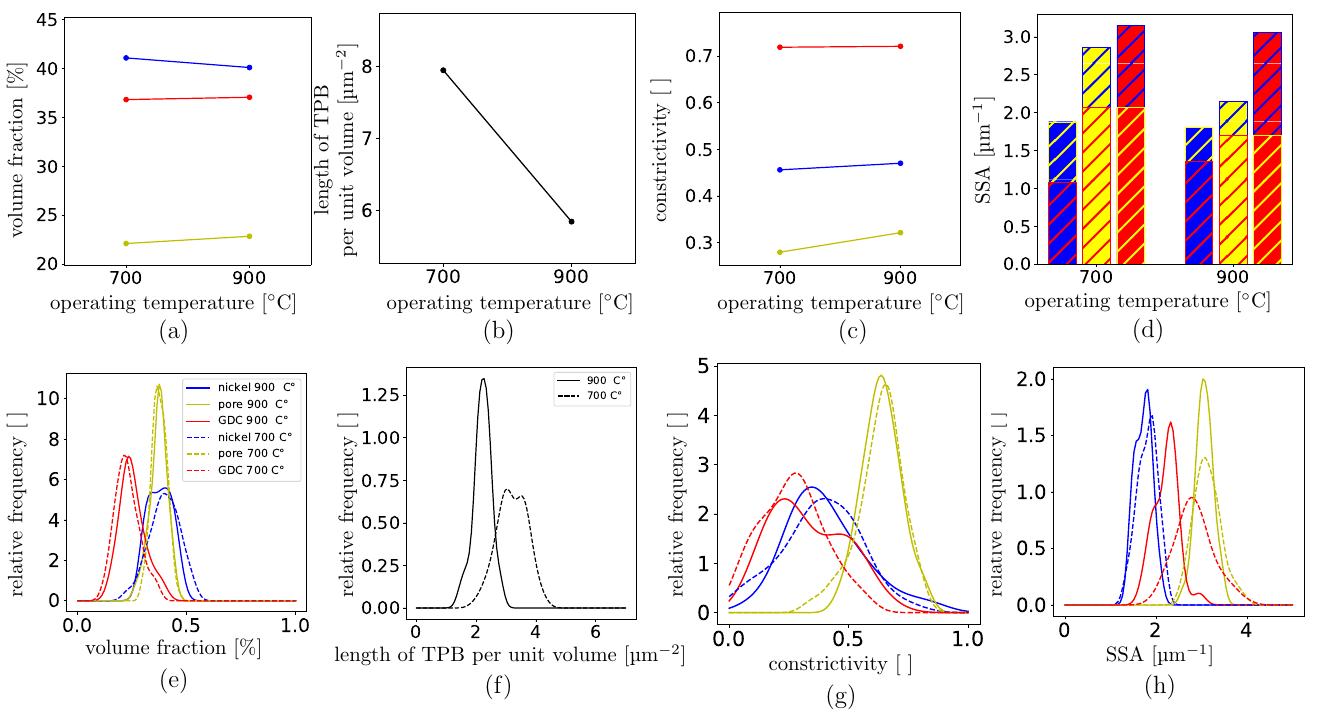}
\end {center}    \caption{Geometrical descriptors  of the three phases nickel (blue),  \cgo (red) and pore space (yellow) for different operating temperatures. Upper row:  volume fraction (a),  specific length of  TPB (b),  constrictivity (c), as well as  SSA (d), which is divided into the contributions from the different interfaces through the hatching, is shown. Lower row: probability densities of local descriptors based on non-overlapping cubic cutouts of size \SI{2.5}{\micro \meter}.}
    \label{fig: temp_global}
\end{figure}

The two samples are compared considering the geometrical descriptors introduced in Section~\ref{sec: geom_desc}.
The upper row of Figure~\ref{fig: temp_global} shows that there are only small differences in  constrictivity for the different operating temperatures. Thus, the higher operating temperature does not lead to more bottleneck effects in the transport property of the three phases.
However, the higher operating temperature significantly reduces the SSAs of the pore space and the associated interfaces, while the volume fraction changes only slightly. We therefore observe a coarsening of the pore space.  Remember that the chemical reaction requires the GDC-pore interface or the TPB. The coarsening of the pore space combined with a significant decrease of the specific length of TPB indicates a worse electrochemical performance of the \SI{900}{\celsius} sample compared to the operating temperature of \SI{700}{\celsius}.

The (global) geometrical descriptors considered in the upper row of Figure~\ref{fig: temp_global} have also  been  computed for non-overlapping cutouts with a size of  \SI{2.5}{\micro\meter}.
The resulting probability densities of local descriptors are shown in the lower row of Figure~\ref{fig: temp_global}. 
Differences between these densities for the two operating temperatures are especially recognizable in  SSA  of the \cgo phase and the pore space, while the nickel phase show only small local differences. In particular, the SSA distribution of \cgo is shifted to the left for a higher operating temperature, indicating a coarsening of the \cgo phase. Furthermore, there are  local differences in the specific length of TPB between the two operating temperatures, whereby the  distribution of the local  specific length of  TPB is broader and its mean is shifted to the right-hand side for the operating temperature of \SI{700}{\celsius}, compared to \SI{900}{\celsius}. Thus, as in the global analysis a shortened TPB is visible for a higher operation temperature. This indicates again a worse electrochemical performance of the \SI{900}{\celsius} sample due to less active reaction sites. Note that similar effects in the \cgo phase have already been reported in \cite{liu.2024}, where an operating temperature of \SI{900}{\celsius} causes a sintering effect that results in a coarsening of the particles of the GDC framework.\\

\begin{figure}[h!]
    \centering
    \includegraphics[width = 0.6\textwidth]{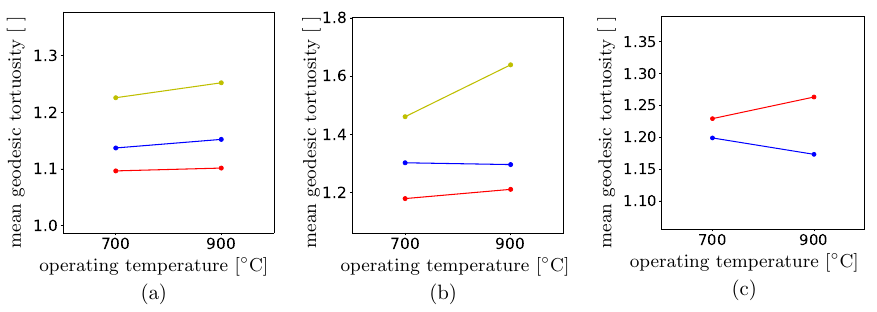}
    \caption{Mean geodesic tortuosity (a), mean geodesic tortuosity of paths starting from TPB (b), and dilated mean geodesic tortuosity with $r$ = \SI{0.1}{\micro \meter} (c), for the three phases nickel (blue), \cgo (red) and pores (yellow),   for different operating temperatures.}
    \label{fig: temp_tortuosity}
\end{figure}

In Figure~\ref{fig: temp_tortuosity}, the mean geodesic tortuosity, the mean geodesic tortuosity of paths starting from  TPB, and the dilated mean geodesic tortuosity with $r =  \SI{0.1}{\micro \meter}$  are shown for all three phases in the respective transport direction. The  mean geodesic tortuosities for the nickel and \cgo phases change only slightly when changing the operating temperature, while the mean geodesic tortuosity of paths starting from the TPB  increases for the pore space. This means that the transport properties of the pore space have worsened, while the transport properties of the \cgo phase and nickel phase remain almost unchanged when the operating temperature is increased to \SI{900}{\celsius}.  Note that the dilated mean geodesic tortuosity of  pore space could not be computed, because the paths within the pore space are too narrow, which indicates a limited transport capacity of the pore space.

\begin{figure}[htb]
    \centering
    \includegraphics[width = 0.8\textwidth]{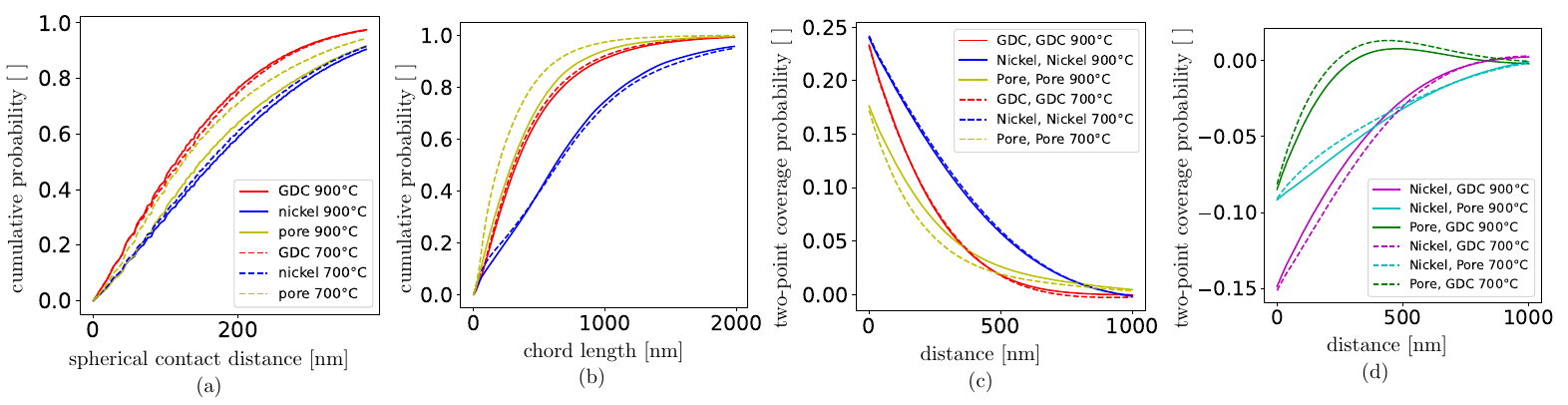}
    \caption{Cumulative distribution functions of spherical contact distance (a) and   chord length (b) for the three phases nickel (blue), \cgo (red)  and pores (yellow), as well as    centered two-point coverage probability functions for each of these phases (c), and for  combinations of different phases  (d), for an operating temperature of \SI{700}{\celsius} and  \SI{900}{\celsius}, respectively.}
    \label{fig: temp_chord}
\end{figure}

In Figure~\ref{fig: temp_chord}, the cumulative distribution functions of  spherical contact distance  and  chord length are shown for the three phases nickel, GDC  and pores, as well as the centered two-point coverage probability functions for each of these phases, and for combinations of different phases, for an operating temperature of \SI{700}{\celsius} and \SI{900}{\celsius}, respectively. 

Again, like in  Figure~\ref{fig: temp_tortuosity}, for the nickel and \cgo   phases only slight changes of the  descriptors considered in Figures~\ref{fig: temp_chord}a and~\ref{fig: temp_chord}b
can be observed when increasing the operating temperature from  \SI{700}{\celsius} to \SI{900}{\celsius}, while the differences for the pore space are more pronounced, for both the spherical contact distances and the chord lengths. More precisely,  for the pore space, an operating temperature of \SI{900}{\celsius} leads to larger contact distances and longer chord lengths than an operating temperature of \SI{700}{\celsius}, which can be interpreted as  coarsening of  pore space.
Furthermore, while the differences between the centered two-point coverage probability functions for the nickel and \cgo phases are negligible when comparing these functions for the operating temperatures of \SI{700}{\celsius} and  \SI{900}{\celsius}, the centered two-point coverage probability function of the pore space decreases more slowly towards zero for \SI{900}{\celsius} compared to  its behavior for \SI{700}{\celsius}, see Figure~\ref{fig: temp_chord} c. With regard to centered two-point coverage probability functions for  combinations of  different phases, it is noticeable that one observes a slower increase towards zero for \SI{900}{\celsius} than for \SI{700}{\celsius}  in case of both phase combinations including the pore space, see Figure~\ref{fig: temp_chord}d. These observations again indicate a coarsening of the pore space where such a coarsening can lead to a reduction of active reaction sites. However, these changes indicate a better transport in the pore space, and hence, the paths have widened at least slightly. Note that the rather limited changes observed in the nickel phase are expected due to the relatively coarse Ni skeleton of the infiltrated cells.

\subsection{Influence of annealing time}
\label{sec: time}

\begin{figure}[ht]
    \includegraphics[width = 0.9 \textwidth]{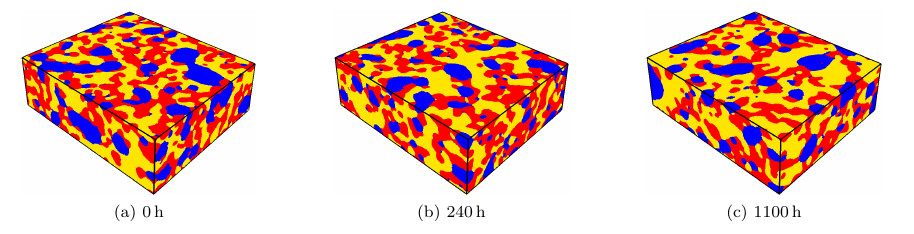}
    \caption{ 3D morphology of SOFC anodes, consisting of nickel (blue), GDC (red) and pore space (yellow), which have been
    manufactured by means of powder technology and  aged with an operating temperature of $\SI{900}{\celsius}$ and  annealing times of $\SI{0}{\hour}$ (a), $\SI{240} {\hour}$ (b) and $\SI{1100}{\hour}$ (c),  where  imaging was  done with a voxel size of $\SI{50}{\nano \meter}$.}
    \label{fig: time_sample}
\end{figure}

To investigate the influence of  annealing time on the 3D morphology of SOFC anodes, we consider the samples manufactured by means of powder technology and a reduction temperature of \SI{800}{\celsius}, which are aged with an operating temperature of \SI{900}{\celsius} and imaged with  a voxel size of \SI{50}{\nano \meter}. More precisely, we compare samples A, B and C, \ie the 3D morphology in pristine state and after  annealing times of \SI{240}{\hour} and  \SI{1100}{\hour}, respectively. The 3D morphology of the three samples A, B and C is visualized in Figure~\ref{fig: time_sample}.
As already seen from the visualization of the 3D microstructure in Figure~\ref{fig: time_sample}, the nickel phase is poorly connected. More precisely, it turns out that the proportion of perculating phase remains below 5 \% for all three samples and the initial polarization resistance measured at \SI{600}{\celsius} with 50 \% humidity is \SI{1.46}{\ohm cm^2}. This high polarization resistance is partially attributed to the high reduction temperature used during electrode formation. As shown in \cite{liu.2024}, a lower reduction temperature decreases the polarization resistance but results also in almost no percolation of the nickel phase. In \cite{kullmann2025impact}, measured ionic and electronic conductivity of GDC is reported. It is found that GDC has inherently poorer ionic conductivity, which limits the electrochemical performance.

\begin{figure}[H]
    \centering
    \includegraphics[width = 0.75\textwidth]{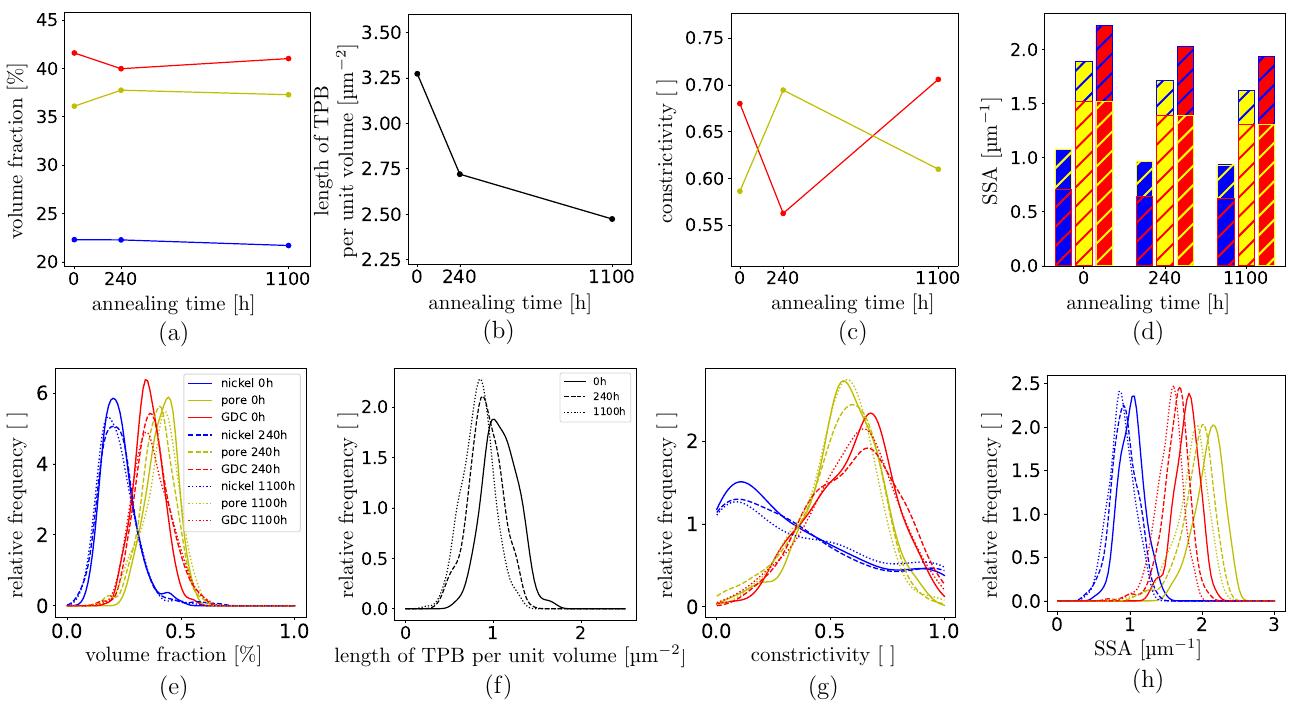}
    \caption{ Geometrical descriptors  of the three phases nickel (blue),  \cgo (red) and pore space (yellow) for different annealing times. Upper row:  volume fraction (a),  specific length of  TPB (b), constrictivity (c), as well as  SSA (d), which is split into the contributions from the   different interfaces through the hatching.  Lower row:  probability densities of local descriptors based on non-overlapping cubic cutouts of size \SI{2.5}{\micro \meter}.}
    \label{fig: time_global}
\end{figure}

The influence of annealing time is  investigated by means of various global  geometrical descriptors, which have been computed on image data of the whole samples, as well as by means of local descriptors, see Figure~\ref{fig: time_global}.

It turned out that there are only negligible changes in the volume fraction for all three phases, see Figure~\ref{fig: time_global}a. But, on the other hand, it is noticeable that the specific length of  TPB significantly decreases with increasing annealing time, whereas the impact of annealing time on constrictivity of \cgo and pore space is less clear, see Figures~\ref{fig: time_global}b and~\ref{fig: time_global}c. Thus, the active reaction sites decrease, but no increase in transport bottleneck effects is indicated in \cgo and pore space. However,  for each operation time considered in this paper, the nickel phase is poorly connected, leading to a constrictivity of zero. Note that a constrictivity of zero does not imply that the SOFC anode is not operational. Furthermore, the SSAs  of nickel and \cgo  as well as of the pore space are  decreasing with increasing annealing time,  while the volume fractions remain almost unchanged, see Figures~\ref{fig: time_global}a and~\ref{fig: time_global}d,  which means that the entire 3D morphology of the SOFC anodes is coarsening. This coarsening in combination with a significant decrease of the specific length of TPB indicates a worse electrochemical performance for longer annealing times.

In the lower row of Figure \ref{fig: time_global}, the  probability densities of local geometrical descriptors are shown, which have been computed for non-overlapping cubic cutouts of size \SI{2.5}{\micro \meter}, where slight differences can be observed for all descriptors of the three phases and for all  annealing times. 
For example, the  probability density of the local specific length of  TPB  is shifted to the left 
for longer annealing times, see  Figures~\ref{fig: time_global}f. This means that the local specific lengths of TPB tend to become shorter locally with increasing annealing time, which again is a reduction of the active reaction sites. It is also remarkable that the nickel phase is locally better connected than globally, which is indicated by the fact that there is a subset of the unit interval $[0,1]$ where the values of the probability density of local constrictivity of the nickel phase are  greater than zero, see  Figure~\ref{fig: time_global}g. Thus, on a local scale, electronic transport  within the nickel phase is more efficient even if big bottleneck effects exist.  \\

\begin{figure}[H]
    \centering
    \includegraphics[width = 0.55\textwidth]{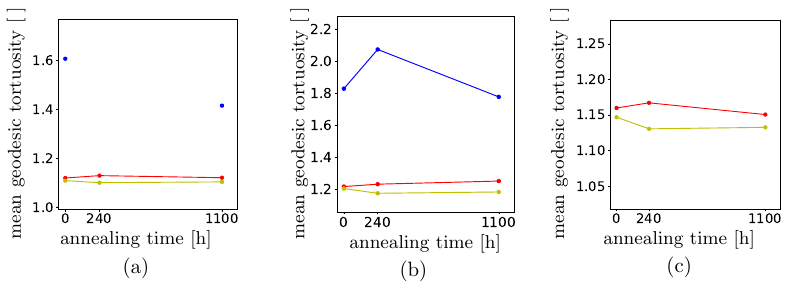}
    \caption{Mean geodesic tortuosity (a), mean geodesic tortuosity of paths starting from TPB (b),  dilated mean geodesic tortuosity with $r = \SI{0.1}{\micro \meter}$ (c), for nickel  (blue),  \cgo  (red),  and pores  (yellow).  
    }
    \label{fig: time_tortuosity}
\end{figure}

To examine the transport capability within the three phases further, the three different types of geodesic tortuosity explained in Section~\ref{sec: geom_desc} are considered. While the nickel phase is too disconnected to determine a mean geodesic tortuosity for the annealing time of \SI{240}{\hour}, the corresponding mean geodesic tortuosity of paths starting from the TPB can be computed, but has a high value, see Figures~\ref{fig: time_tortuosity}a and~\ref{fig: time_tortuosity}b. Moreover, the dilated mean geodesic tortuosity with $r = \SI{0.1}{\micro \meter}$ cannot be computed for the nickel phase in case of all annealing times, see Figure~\ref{fig: time_tortuosity}c, which indicates that the existing paths in the nickel phase are rather narrow. Thus, the poor connection and the narrow paths in the nickel phase indicate that the electric transport properties of this phase are rather poor. In contrast, the \cgo phase and the pore space have a relatively low mean geodesic tortuosity values with regard to all three types of geodesic tortuosity indicating efficient transport in these phases.

\begin{figure}[h!]
    \centering
    \includegraphics[width = 0.8\textwidth]{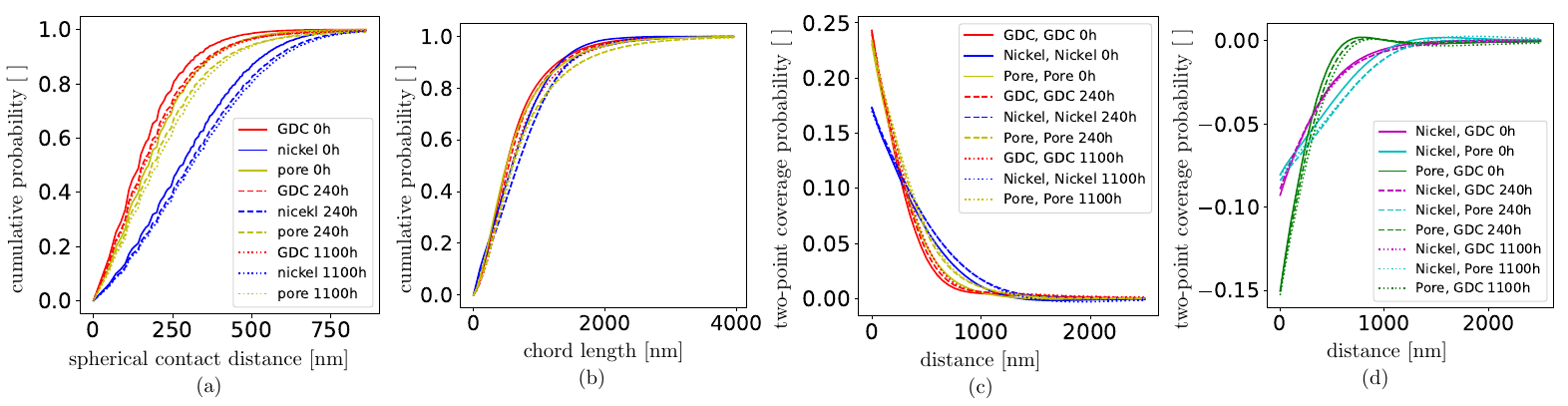}
    \caption{Cumulative distribution functions of  spherical contact distance (a) and  chord length (b) for  nickel (blue), GDC (red) and pores (yellow), as well as centered two-point coverage probability functions for each of these phases (c)  and  for combinations of different phases (d).}
    \label{fig: time_chord}
\end{figure}

Furthermore, the distributions of spherical contact distances  and  chord lengths have been analyzed to investigate the impact of annealing time on the 3D morphology of SOFC anodes. For all three phases, the spherical contact distances tend to  increase with increasing annealing time,  see Figure~\ref{fig: time_chord}a. Similarly, in comparison to the pristine state, the chord lengths get longer during operation, especially for the nickel phase,  see Figure~\ref{fig: time_chord}b.  Thus, also these observations  indicate a coarsening of the phases in the course of fuel cell operation.

Examining the centered  two-point coverage probability functions of the individual phases shows that the values of these functions  increase  with increasing annealing time,  see Figure~\ref{fig: time_chord}c.  Interestingly, this increase is more pronounced during shorter annealing times indicating more pronounced structural changes at the early stage of operation compared to the later one. Examining the centered two-point coverage probability functions of two different phases, it turns out that they are shrinking with increasing annealing time,  see Figure~\ref{fig: time_chord}d. Thus, again a coarsening of the 3D anode morphology with increasing annealing time is indicated, which directly correlates to shorter TPB as well as a smaller GDC-pore interface. Both properties are required for the chemical reaction.

\subsection{Influence of manufacturing process}

\begin{figure}[ht!]
    \centering
    \includegraphics[width = 0.7\textwidth]{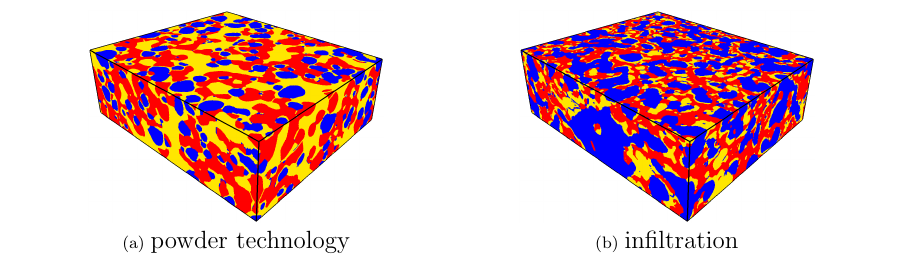}
    \caption{3D morphology of pristine SOFC anodes prepared by   powder technology (a) and infiltration (b), 
    comprising nickel (blue), \cgo (red),  and pores (yellow).}
    \label{fig: manu_sample}
\end{figure}

Last but not least, we compare the 3D morphologies of pristine SOFC anodes prepared by means of two different manufacturing processes, \ie powder technology and infiltration, where samples E and G are used, \ie 3D image data of pristine samples with a voxel size of \SI{20}{\nano \meter} and manufactured with a reduction temperature of \SI{650}{\celsius}, see  Figure~\ref{fig: manu_sample}.

\begin{figure}[H]
    \centering
    \includegraphics[width = 0.7\textwidth]{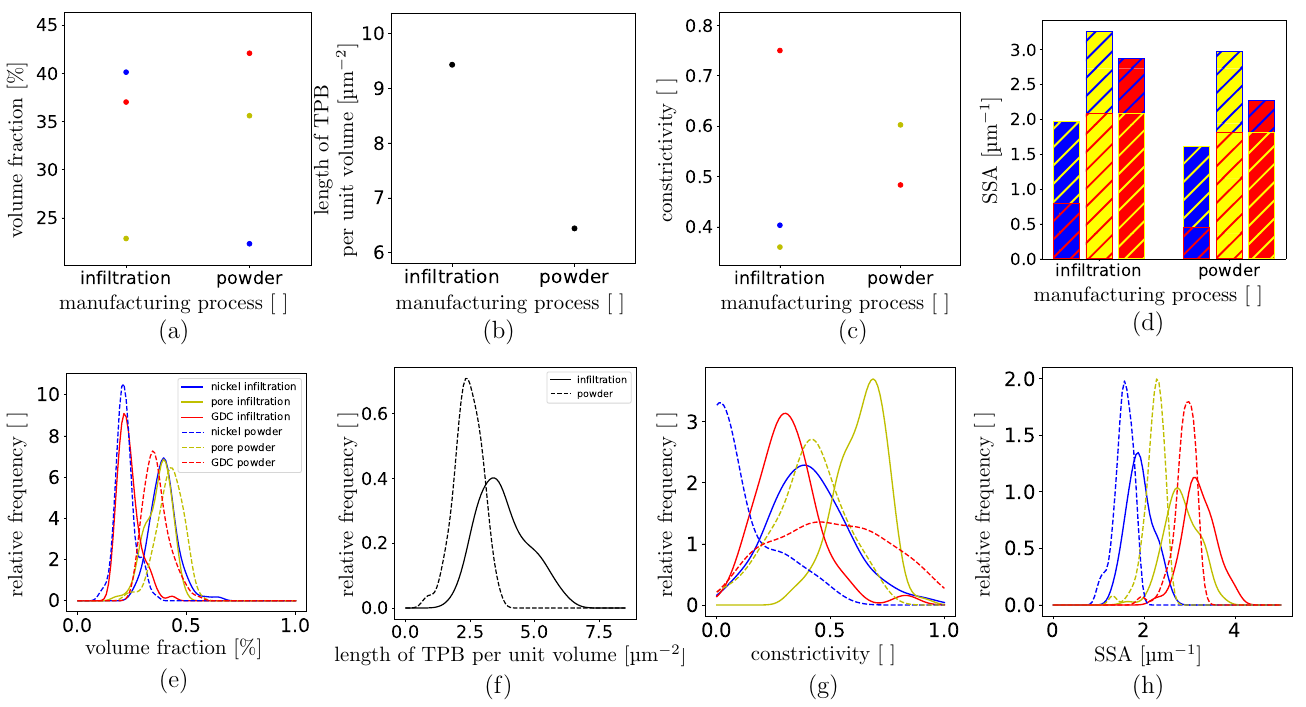}
    \caption{Geometrical descriptors  of the three phases nickel (blue), \cgo  (red)  and pore space (yellow) for different manufacturing processes. Upper row:  volume fraction (a),  specific length of  TPB (b), constrictivity (c) as well as SSA (d), which is split into the contributions from the different interfaces through the hatching.  Lower row: probability densities
of  local  descriptors based on non-overlapping cutouts of size \SI{2.5}{\micro \meter}.}
    \label{fig: manu_global}
\end{figure}

First, to investigate the influence of the manufacturing process on the 3D morphology of the anodes, geometrical descriptors are computed on the whole anode  samples. It is striking that the volume fractions of the three phases (nickel, \cgo and  pores) differ significantly for the two samples prepared by the different manufacturing processes. For example, the volume fraction of nickel phase in the SOFC anode manufactured by infiltration is roughly \SI{40}{\percent}, while it is halved for the SOFC anode manufactured by means of the powder technology,  see  Figure~\ref{fig: manu_global}a. Note that the differences with regard to volume fractions are not only caused by the different manufacturing processes, but also due to different material compositions, which have been used for the different manufacturing processes. This has to be taken into account with regard to the results of the subsequent geometrical 
descriptors. Therefore, a direct comparison between the two manufacturing processes is not possible with the given samples. Nevertheless, we would like to examine the existing samples further on the basis of their geometrical descriptors.

As shown in Figure~\ref{fig: manu_global}b,
the anode manufactured by means of the infiltration technology has a significantly larger specific length of TPB than the anode manufactured by means of the powder technology. However, one should be aware that this difference is not only caused by the manufacturing process. The TPB boundary is also influenced by the different phase fractions. Besides this, the nickel phase obtained by powder technology is very disconnected  such that its constrictivity is equal to zero, while  infiltration with a higher nickel amount leads to a (small) positive constrictivity value of the nickel phase, see    Figure~\ref{fig: manu_global}c. 
Furthermore, there are differences in the SSA between the two samples, see Figure~\ref{fig: manu_global}d. In particular, the fraction of the nickel-\cgo interface and the SSAs of all three phases are smaller for the anode manufactured with the powder technology than for the anode manufactured with the infiltration technology. 
It should be particularly emphasized that the powder technology does not yield a larger GDC surface area despite the higher GDC amount. This indicates that the infiltration technology results in a finer GDC structure. The same applies for the pore space. This leads also to a higher GDC-pore interface for the sample manufactured by means of infiltration technology. Altogether,  more active reaction sites and a more desirable 3D morphology is achieved by the infiltration technology.
For non-overlapping cutouts of size \SI{2.5}{\micro \meter}, the corresponding probability densities of  local geometrical descriptors are shown in the lower row of Figure~\ref{fig: manu_global}. Here, too, it is clearly visible that the anode with the higher nickel fraction and manufactured with infiltration technology, has a longer local specific length of  TPB than the anode with lower  nickel fraction and manufactured with powder technology. In addition, the local SSAs of all three phases of the infiltration-manufactured anodes have a wider and more right-shifted distribution than the anode manufactured with powder technology, indicating  larger local surfaces in the infiltration-manufactured anode. Thus, the anode manufactured with infiltration technology possesses more active reaction sites.
\\

\begin{figure}[ht!]
    \centering
    \includegraphics[width = 0.6\textwidth]{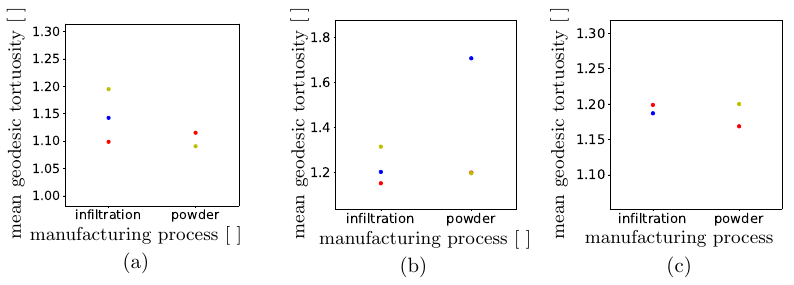}
    \caption{Mean geodesic tortuosity (a), mean geodesic tortuosity of paths starting from TPB (b) and dilated mean geodesic tortuosity with $r$ = \SI{0.1}{\micro \meter} (c), for nickel (blue), \cgo  (red) and pores (yellow).}
    \label{fig: manu_tortuosity}
\end{figure}

In Figure~\ref{fig: manu_tortuosity},  results are shown, which have been obtained for the three different types of mean geodesic tortuosity.  The mean geodesic tortuosity and the dilated mean geodesic tortuosity can not be computed for the nickel phase of  anodes manufactured by  powder technology, due to the poor connection within the nickel phase which is already discussed in Section \ref{sec: time}. Nevertheless, the anode has a small tortuosity value for the pore space and CGO, which indicates good transport properties in these phases.
Note that the dilated mean geodesic tortuosity of  pore space can not be computed for  SOFC anodes manufactured by  infiltration, because the paths within their pore space were  very narrow. In general, however, the anode manufactured by infiltration has  relatively low tortuosity values for all three phases, which indicates good transport properties in all phases.

\begin{figure}[h]
    \centering
    \includegraphics[width = 0.85\textwidth]{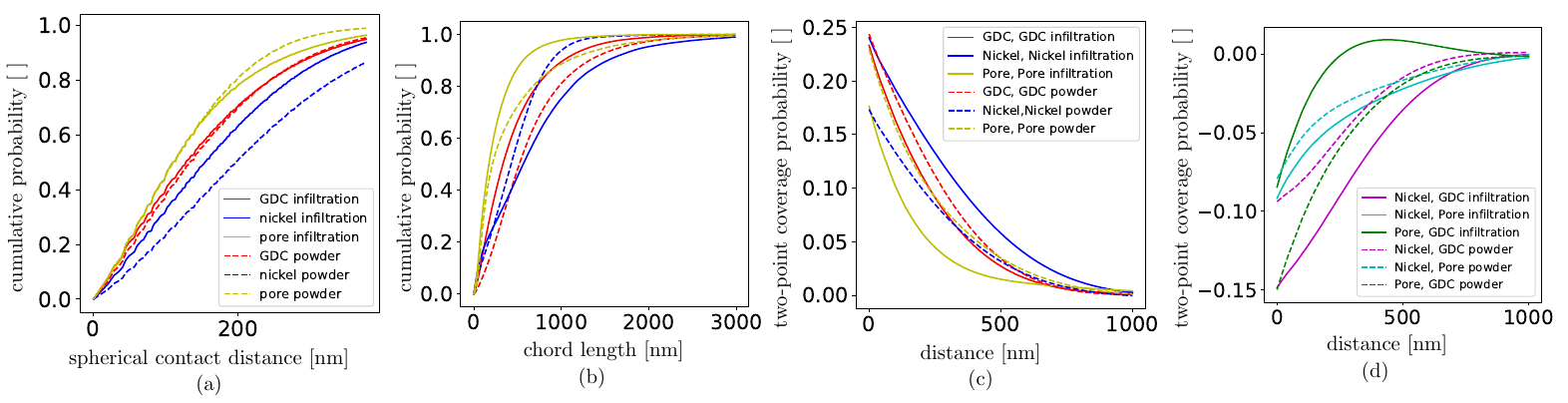}
    \caption{Cumulative distribution functions of spherical contact distance (a) and chord length (b) for nickel (blue),
GDC (red) and pores (yellow), as well as centered two-point coverage probability functions for each of these phases
(c) and for combinations of different phases (d).}
    \label{fig: manu_chord}
\end{figure}

Last but not least, the  spherical contact distances and chord lengths of the three phases, \ie nickel, GDC and pores, have been analyzed. For the nickel phase, the cumulative distribution function values of spherical contact distances are lower for the sample manufactured by means of powder technology and a lower nickel fraction, see Figure~\ref{fig: manu_chord}a, which indicates  a coarser nickel phase structure for the anode manufactured with powder technology. Additionally, the anode with higher nickel fraction and manufactured by infiltration has lower cumulative probabilities for  chord lengths in the nickel phase than the anode manufactured by means of powder technology, indicating a  more elongated nickel phase structure, see Figure~\ref{fig: manu_chord}b. For  GDC  and pores,  only minor differences are observed in the spherical contact distance distributions for the two different anodes. However, the anode manufactured with infiltration and with a higher nickel amount has higher cumulative probabilities of  chord lengths for both phases, GDC  and pores, compared to the anode manufactured by means of powder technology, which indicates a finer pore space and \cgo phase. 

The two-point coverage probabilities of both, \cgo and pores, are larger for the anode manufactured by means the powder technology than for the anode manufactured with infiltration, while the opposite can be observed for the nickel phase, see Figure~\ref{fig: manu_chord}c. However, this observation can be caused by the different material composition, as the amount of GDC and pores is significantly greater and the nickel amount is significantly smaller in the anode manufactured by means of the powder technology. Moreover, the centered two-point coverage probabilities of the pore-\cgo combination are significantly larger for the anode manufactured by means of the infiltration technology compared to anode manufactured by the powder technology. For the remaining phase combinations, the anode manufactured by means of powder technology has larger centered two-point coverage probabilities compared to the anode manufactured by the infiltration process, see Figure~\ref{fig: manu_chord}d.

\section{Conclusion}

In the present paper,  FIB-SEM image data have been used to analyze the 3D morphology of seven SOFC anodes, comprising of nickel, GDC and pore space. For this purpose, various geometric descriptors such as volume fraction, SSA,  specific length of  TPB and  three different types of mean geodesic tortuosity have been computed. Besides the computation of these descriptors for each entire sample, their local heterogeneity is quantified by determining the distributions of local geometrical descriptors based on non-overlapping cutouts. This data enables the quantitative examination of the effect of various influencing factors  (operating temperature, annealing time and  manufacturing process) on the resulting 3D morphology of SOFC anodes. It is shown that all three  influencing factors mentioned above affect the  anode morphology. In particular, the higher operating temperature of \SI{900}{\celsius} leads to a coarsening of the pore space. Similar effects can be observed with respect to annealing time, where it is noticeable that the specific length of TPB significantly decreases with increasing annealing time. Moreover, pronounced differences between both considered manufacturing processes have been observed, suggesting that infiltration leads to a finer anode structure. These structural differences can also be caused by the different material compositions, which have been used for the different manufacturing processes. \\

In a forthcoming study, the structural changes of SOFC anodes considered in the present paper will be modeled stochastically. In particular,  a suitable spatial stochastic  model will be determined, whose parameters will be calibrated to 3D image data of SOFC anodes for various annealing times. Furthermore, by interpolation within the parameter space, it will be possible to carry out fast predictive simulations  of SOFC anodes for annealing times for which no image data is available. In this way, an efficient method will be established in order to provide the geometry input for spatially resolved numerical simulations of effective macroscopic properties. This will allow for efficient virtual testing of SOFC anodes, including the derivation of quantitative process-structure-property relationships.  

\section*{Acknowledgement}

The authors would like to thank the Federal Ministry of Education and Research (BMBF) for financial support within the priority program ``Mathematics for Innovations'' (grant number 05M2022) and the project ``WirLebenSOFC'' (grant numbers 03SF0622E and 03SF0622B). BN acknowledges support provided by the Helmholtz Association within programme MTET, no. 38.04.04. 

\bibliography{Sec_6_Bibliography}{}
\bibliographystyle{elsarticle-num}

\end{document}

%% file: header.tex
\usepackage[utf8]{inputenc}
\usepackage{cite}
\usepackage{amsmath}
\usepackage{amssymb}
\usepackage{graphicx}
\usepackage{subcaption}
\usepackage{dsfont}  
\usepackage{authblk}
\usepackage{hyperref}
\usepackage{siunitx}
\usepackage{xcolor}
\usepackage{caption}
\usepackage[utf8]{inputenc}
\usepackage{textcomp}
\usepackage{lineno}
\usepackage{float}
\usepackage{booktabs}
\usepackage{adjustbox}
\usepackage{tikz}
\usetikzlibrary{shapes, arrows, positioning,decorations.markings}
\tikzstyle{vecArrow} = [thick, decoration={markings,mark=at position
	1 with {\arrow[semithick]{open triangle 60}}},
double distance=1.4pt, shorten >= 5.5pt,
preaction = {decorate},
postaction = {draw,line width=1.4pt, white,shorten >= 4.5pt}]
\tikzstyle{innerWhite} = [semithick, white,line width=1.4pt, shorten >= 4.5pt]
\hypersetup{hidelinks}
\usepackage[tmargin=1in,bmargin=1in,lmargin=0.75in,rmargin=0.75in]{geometry}

%% file: makros.tex
\newcommand{\R}{\ensuremath{\mathbb{R}}}

\newcommand{\cgo}{GDC }
\newcommand{\ie}{\textit{i.e., }}

\providecommand{\keywords}[1]
{
	{
	\textbf{\textit{Keywords---}} #1
	}
}

\newcommand\blfootnote[1]{%
	\begingroup
	\renewcommand\thefootnote{}\footnote{#1}%
	\addtocounter{footnote}{-1}%
	\endgroup
}

%% file: abstract.tex
\begin{abstract}
\noindent

Solid oxide fuel cells (SOFCs) 
are becoming increasingly important due to their high electrical efficiency, the flexible choice of fuels and relatively low emissions of pollutants. However, the increasingly growing demands for electrochemical devices require further performance improvements as for example by reducing degradation effects. Since it is well known that the 3D electrode morphology, which is significantly influenced by the underlying manufacturing process, has a profound impact on the resulting performance, a deeper understanding for the structural changes caused by modifications of the manufacturing process or degradation phenomena is desirable. In the present paper, we investigate the influence of the annealing time and the operating temperature on the 3D morphology of SOFC anodes using 3D image data obtained by focused-ion beam scanning electron microscopy, 
which is segmented into gadolinium-doped ceria, 
nickel and pore space. In addition, structural differences caused by manufacturing the anode via infiltration or powder technology, respectively, are analyzed quantitatively by means of various geometrical descriptors such as  specific surface area, mean geodesic tortuosity, and constrictivity. The computation of these descriptors from 3D image data is carried out both globally as well as locally to quantify the heterogeneity of the anode structure. 

\end{abstract}
\keywords{solid oxide fuel cell; 3D morphology;  FIB-SEM; anode;  degradation; aging; statistical image analysis}